\documentstyle[12pt,epsf]{article}
\input epsf.tex

\begin{document}

\thispagestyle{empty}

\vspace{24pt}
\begin{center}

{\large\bf MEVSIM: A Monte Carlo Event Generator for STAR}\\

\vspace{24pt}

R. L. Ray \\

\vspace{12pt}

{\sl Department of Physics}\\
{\sl The University of Texas at Austin, Austin, Texas 78712}\\

\vspace{24pt}

R. S. Longacre \\

\vspace{12pt}

{\sl Physics Department, Brookhaven National Laboratory, Upton, NY  11973}\\

\begin{abstract}
A fast, simple to use Monte Carlo based event generator is presented which is
intended to facilitate simulation studies
and the development of 
analysis software for the Solenoidal Tracker at RHIC (Relativistic Heavy Ion
Collider) (STAR) experiment at the Brookhaven National Laboratory (BNL). 
This new event generator provides a fast,
convenient means for producing large numbers of uncorrelated A+A collision
events which can be used
for a variety of applications in STAR, including quality assurance
evaluation of event reconstruction software, determination of detector
acceptances and tracking efficiencies, physics analysis of event-by-event
global variables, studies of strange, rare and exotic particle reconstruction,
and so on.  The user may select the number of events, the particle
types, the multiplicities, the one-body momentum space distributions and the
detector acceptance ranges.  The various algorithms used in the code and
its capabilities are explained.  Additional user information is also 
discussed.  The computer code implementation is called {\sc mevsim}.

\end{abstract}

\end{center}

\section{Introduction}
\label{section:introduction}

When carrying out STAR detector simulations and physics analysis studies
one often needs many events of a similar type that can either be readily
selected or tightly controlled by the user to meet the specific needs of
the simulation project.  For example, in evaluating the performance of the
event reconstruction software it is beneficial to use events which have
the same particle multiplicities and distributions from event-to-event in
order to help identify anomalous changes in the tracking performance. 
Events with flat
distributions in rapidity facilitate the determination of detector acceptances
and tracking efficiencies.  Events with enhanced numbers of multi-strange
particles and other rare particles facilitate the study of the corresponding
reconstruction and physics
analysis software.  When using standard events generators
such as {\sc hijing}, {\sc venus}, {\sc rqmd}, {\sc vni}, etc. the user
has little control over the simulated
particle production other than to select the
beam energy and colliding nuclei,
impact parameter, and perhaps the settings of some physics
process control switches.  Furthermore, most of these codes
require significant amounts of cpu-time.  For many STAR simulation studies
these sophisticated models are not really necessary and unless the user
is interested in the inherent particle correlations which a given model
may produce, a simple Monte Carlo particle generator will generally suffice.

In response to this need we have developed a standalone Fortran~77 computer
code called {\sc mevsim}.
The code is designed to provide a fast,
convenient means for producing large numbers of uncorrelated A+A collision
events which can be used for a variety of
applications in STAR, including quality assurance
evaluation of event reconstruction software, determination of detector
acceptances and tracking efficiencies, physics analysis of event-by-event
global variables, studies of strange and rare particle reconstruction,
and so on.  The user may select the detector acceptance ranges,
the number of events, the particle
types, the multiplicities, and the one-body momentum space distributions with
respect to transverse momentum ($p_T$), rapidity ($y$) and azimuthal
angle ($\phi$) from a menu of models.  

The particle multiplicities can either be fixed for each event or allowed
to randomly vary from event-to-event according to Poisson statistics.  The
parameters of the one-body momentum space distributions ({\it e.g.}
slope parameter, rapidity width, radial expansion velocity) can either be
fixed or randomly varied from event-to-event according to a Gaussian
distribution in order to simulate dynamical fluctuations.  In addition
the user may specify a reaction plane angle and include anisotropic flow.
The overall multiplicities for all particle types in each event can also be
scaled by a common, multiplicative factor in order to simulate trigger
fluctuations.  In addition to the usual list of particle types in
{\sc geant}, the code also includes a number of short lived resonances, several
of which have explicit mass distributions.  These
resonances are not decayed.
Finally, the events produced are assumed to
be in the A+A center-of-momentum frame which for symmetric beam species at
RHIC coincides with the lab frame.

In the following sections we describe the random sampling methods used
in the code, the one-body momentum space distribution models, the
acceptance cuts for application to STAR, the implementation of resonances,
trigger fluctuation simulations, reaction plane and anisotropic
flow simulations, and the standalone code and related files.

\section{Random Sampling Methods}
\label{section:random}

The method used in the code to randomly sample a one-dimensional probability
distribution is described first, followed by an explanation of the random
selection - random rejection algorithm which is used to populate momentum
space according to the specified one-body distributions.

In order to randomly sample a one-dimensional
probability distribution $P(x)$ throughout the domain $x_1 \leq x \leq x_2$
one first obtains the integral,
\begin{equation}
F(x) = \int_{x_1}^{x} P(x^{\prime}) dx^{\prime}~,
\end{equation}
normalized such that $F(x_2) = 1$. 
By finding the value of $x$ such that $F(x) = f$,
where $f$ is a randomly selected number
between 0 and 1, a random collection of $x$ values can be obtained.
A histogram of these values of $x$ will be statistically consistent with
the probability distribution $P(x)$.
This method is much faster than the usual random selection - random
rejection method since cpu-time is not wasted on trial
values of $x$ which end up being discarded. 
In the code the above is implemented
by integrating $P(x)$ on a fixed mesh, randomly selecting $f$ and interpolating
the inverse $F$ function on an uneven mesh using a Lagrange interpolation
routine.

The preceding method is used to sample the reaction plane angle, the overall
multiplicity rescaling factor, the particle multiplicities, the parameters
of the one-body momentum space distribution model, the anisotropic flow
parameters, and the resonance masses.  Generally these follow Gaussian
probability distributions,
\begin{equation}
P(x,x_{\rm mean},\sigma) = N {\rm exp}\left[ -(x - x_{\rm mean})^2/2\sigma^2
\right]~,
\end{equation}
where $N$ is a normalization constant and
the user specifies the mean value, $\sigma$ and number of $\sigma$s
(domain) over which $x$ is to be sampled.  The particle multiplicities
follow a Poisson distribution,
\begin{equation}
P(x,x_{\rm mean}) = \frac{x_{\rm mean}^x e^{-x_{\rm mean}}}{x!}
\end{equation}
with mean value $x_{\rm mean}$ and standard deviation $\sqrt{x_{\rm mean}}$.
The user specifies the mean multiplicity and the ($\pm$) number of standard
deviations about the mean over which to sample.  The resonance masses follow
approximate Breit-Wigner distributions which are listed in
Sec.~\ref{section:resonances}.
In order for each of these quantities to fluctuate the user must input
positive values for both the standard deviation and the ($\pm$) number of
standard deviations over which to sample.

The momentum space distributions, being two-dimensional in $p_T$ and $y$,
are sampled by the usual random selection - random rejection method. 
Furthermore, since the anisotropic
flow parameters depend on $p_T$ and $y$, it was more convenient
to also sample the azimuthal ($\phi$) distribution by the
random selection - random rejection method,
rather than use the preceding integration-interpolation method.

The arbitrary, two-dimensional function $g(u,v)$, defined in the
domain $u_1 \leq u \leq u_2$ and $v_1 \leq v \leq v_2$, is randomly
sampled as follows.  The maximum value of $g$ ($g_{\rm max}$)
in the domain is determined. 
Values of $u$ and $v$ in the domain are
randomly selected and the quantity $g(u,v)/g_{\rm max}$ is computed.  A
random number from 0 to 1 is selected and if it is $\leq g(u,v)/g_{\rm max}$
(where $g$ must be non-negative which is the case for particle densities),
then the randomly selected coordinates $u$ and $v$ are retained;
if not, these coordinates 
are rejected and another trial set of $u$ and $v$ is selected.  Such methods
are reliable but can suffer from poor cpu performance if the function
$g(u,v)$ is strongly peaked within its domain.  For RHIC collisions the $p_T$
and $y$ distributions are not expected to have large spikes and we expect this
sampling method to perform well. 
However, as noted in Sec.~\ref{section:flow},
care should be exercised when including $p_T$ and $y$ dependence in the 
anisotropic flow in order to avoid poor cpu performance.

\section{Momentum Distribution Models}
\label{section:models}

For all momentum space distribution models used in the code the
($p_T,y$) and azimuthal ($\phi$) dependences are factored.  The various
models for the ($p_T,y$) dependences are described in this section, while
the $\phi$ dependence models are explained in Sec.~\ref{section:flow}.

A total of six ($p_T,y$) distribution models can be selected; the same
model must be applied to each particle ID type in a run.  The models
include the following:
\begin{enumerate}
\item
Factorized $p_T$ and $y$ dependence \cite{humanic},
\begin{equation}
\frac{d^2N}{dp_T dy} = A p_T {\rm e}^{-m_T/T} {\rm e}^{-y^2/2\sigma_y^2}
\end{equation}
where $A$ is a nomalization constant, $m_T = \sqrt{m^2 + p_T^2}$,
$m$ is the particle mass, $T$ is the
$m_T$ slope parameter or ``temperature,''
and $\sigma_y$ is the rapidity distribution
width.
\item
Non-expanding spherical thermal source (Pratt) \cite{pratt1},
\begin{equation}
\frac{d^2N}{dp_T dy} = A p_T E {\rm e}^{-E/T}
\end{equation}
where $E = m_T {\rm cosh}(y)$ is the total energy.
\item
Non-expanding spherical thermal source (Bertsch) \cite{bertsch},
\begin{equation}
\frac{d^2N}{dp_T dy} = \frac{A p_T E}{{\rm e}^{E/T} - 1}.
\end{equation}
\item
Spherically expanding, thermally equilibrated source (Pratt) \cite{pratt2},
\begin{eqnarray}
\frac{d^2N}{dp_T dy} & = & A p_T E {\rm e}^{-\gamma E/T} \nonumber \\
 & & \times \left[ \frac{\sinh(y_p)}{y_p} + \frac{T}{\gamma E}
\left( \frac{\sinh(y_p)}{y_p} - \cosh(y_p) \right) \right]
\end{eqnarray}
where $y_p = \gamma v p/T$, $\gamma = 1/\sqrt{1 - v^2}$, $v$ is the radial
expansion velocity parameter, and $p = \sqrt{E^2 - m^2}$ is the total momentum.
\item
Arbitrary, factorized $p_T$ and $y$ dependence loaded bin-by-bin,
\begin{equation}
\frac{d^2N}{dp_T dy} = A f(p_T) g(y)
\end{equation}
where $f$ and $g$ are input histograms.
\item
Arbitrary, two-dimensional function of $p_T$ and $y$ loaded bin-by-bin,
\begin{equation}
\frac{d^2N}{dp_T dy} = A h(p_T,y)
\end{equation}
where $h$ is a two-dimensional histogram.
\end{enumerate}

\section{Kinematic Acceptance}
\label{section:acceptance}

The generated particles are confined within specified kinematic
boundaries.  The acceptance cuts are designed to be applicable to
the STAR geometry.  The acceptances are defined by lower and upper values
for $p_T$, pseudorapidity ($\eta$) and azimuthal angle ($\phi$) (between
0$^{\circ}$ and 360$^{\circ}$).

\section{Resonances}
\label{section:resonances}

The available particle types include those listed in {\sc geant} \cite{geant}
plus a number of additional particles and resonances which are listed in
Table~\ref{TableI} along with their assumed mass and width parameters.  For
most of these the widths are sufficiently broad to warrant including the
mass fluctuations explicitly in the generated particle list.  The
mass distribution models assumed in the code
are described in this section.  Since the mass and
width parameters depend somewhat on the model, the parameters listed in
Table~\ref{TableI} may not agree exactly with similar values tabulated
elsewhere, such as in Ref.~\cite{PDG}.  In the code the resonance masses are
sampled after the $p_T$, $y$ and $\phi$ distributions have been determined.
Hence, in the final output particle list the $p_T$, $y$, $\phi$ and mass
distributions will be exactly as specified by the user.  However, because
the masses are changed after $p_T$ and $y$ are fixed, the pseudorapidity
values will change and the pseudorapidity distributions will be altered
by the mass fluctuations.

The following resonance shapes (un-normalized) were assumed:
\begin{enumerate}
\item
$\rho$-meson \cite{rho}
\begin{equation}
W(M) = \frac{q}{(M - M_{\rho})^2 + \Gamma_{\rho}^2/4}~,
\end{equation}
where $q$ is the c.m. momentum in the final $\pi - \pi$ system, and
$M_{\rho}$ and $\Gamma_{\rho}$ are listed in Table~\ref{TableI}.
\item
$\omega$-meson \cite{omega}
\begin{equation}
W(M) = \frac{M_{\omega}^2 \Gamma_{\omega}}{(M_{\omega}^2 - M^2)^2
+ M^2 \Gamma_{\omega}^2(s)} ~,
\end{equation}
where $\Gamma_{\omega}(s) = \Gamma_{\omega} \left(M/M_{\omega} \right)^3$
and $M_{\omega}$ and $\Gamma_{\omega}$ are given in Table~\ref{TableI}.
\item
$\phi$-meson \cite{phi}
\begin{equation}
W(M) = \frac{\Gamma_{\phi}^2/4}{(M - M_{\phi})^2 + \Gamma_{\phi}^2/4}
\left( \frac{q}{q_0} \frac{E_0}{E} \right)^3~,
\end{equation}
where $q$ is the c.m. momentum in the final $K - K$ system,
$E = \sqrt{q^2 + M_K^2}$, $M_K$ is the kaon mass, $q_0$ and $E_0$
are evaluated for $M = M_{\phi}$,
and $M_{\phi}$ and $\Gamma_{\phi}$ are given in Table~\ref{TableI}.
\item
$\Delta$-resonance \cite{Delta}
\begin{eqnarray}
W(M) & = & \frac{4 \pi}{\mu q} \frac{\Gamma(q)^2}
{(M - M_{\Delta})^2 + \Gamma(q)^2/4}~, \\
\Gamma(q) & = & \frac{2 (qR)^3 \Gamma_{\Delta} }{1 + (qR)^2} ~, 
\end{eqnarray}
where $qR = 0.81 q /m_{\pi}$, $q$ is the c.m. momentum in the final $\pi - N$
system, and $\mu$ is the reduced, total energy of the $\pi - N$
system given by,
\begin{equation}
\mu = \frac{\epsilon_{\pi}
\epsilon_{N}}{\epsilon_{\pi} + \epsilon_{N}}~,
\epsilon_{\pi}  = \sqrt{q^2 + m_{\pi}^2}~, 
\epsilon_{N}   = \sqrt{q^2 + M_N^2}~,
\end{equation}
where $m_{\pi}$ and $M_N$ are the pion and nucleon masses, respectively,
and $M_{\Delta}$ and $\Gamma_{\Delta}$ are given in Table~\ref{TableI}.
\item
$K^{\ast}$ resonances \cite{Kstar}
\begin{eqnarray}
W(M) & = & \frac{\Gamma(q)^2 M_{K^{\ast}}^2}{(M^2 - M_{K^{\ast}}^2)^2
+ \Gamma(q)^2 M_{K^{\ast}}^2} ~, \\
\Gamma(q) & = & \frac{2 \Gamma_{K^{\ast}} (q/q_0)^3 }{1 + (q/q_0)^2} ~,
\end{eqnarray}
where $q$ is the c.m. momentum in the final $\pi - K$ system, $q_0$
is evaluated for $M = M_{K^{\ast}}$,
and $M_{K^{\ast}}$ and $\Gamma_{K^{\ast}}$ are given in Table~\ref{TableI}.
\end{enumerate}

\section{Trigger Fluctuations}
\label{section:trigger}

In the STAR A+A event stream, for which the Level 0 triggers are based on
integrated yields in the central trigger barrel (CTB) and multi-wire
chambers (MWC), the actual particle multiplicities
would fluctuate from event-to-event, even if the trigger acceptance settings
could correspond to an ideal $\delta$-function. 
To simulate this effect in the data
stream we included in the code the ability to rescale the multiplicities of
all particle types in the event by a common multiplicative factor, $f$.  This
factor is selected randomly for each event according to a Gaussian 
distribution which the user specifies.

Ideally, the nominal mean multiplicities for each particle type, $N_{\rm pid}$,
should be multiplied by $f$ and the resulting Poisson distribution with
mean value $f N_{\rm pid}$ and standard deviation $\sqrt{f N_{\rm pid}}$
sampled to yield the actual particle multiplicities for the event.
However, it turns out to be much more
straightforward in the code and to have less impact
on the cpu demand, to instead,
sample the nominal multiplicity distributions first
and then
scale the results to get the actual particle multiplicities for the event.
Using this method the particle multiplicity distributions will have the 
correct mean value of $f N_{\rm pid}$ but the incorrect
(non-Poisson) standard deviation
of $f \sqrt{N_{\rm pid}}$ which is in error by the factor $\sqrt{f}$.  It is 
anticipated that $f$ will be within a few percent of 1.0 such that the
actual, non-Poisson multiplicity distributions will only err by a few percent.

\section{Reaction Plane and Anisotropic Flow}
\label{section:flow}

The detection and study of directed flow, elliptical flow and perhaps
higher modes of particle 
flow in non-central A+A collision events at RHIC are of great interest
to the STAR collaboration.
These phenomena can be simulated with this generator.  The user may select
from among four options for the reaction plane:
\begin{enumerate}
\item
Ignore the reaction plane, events are azimuthally invariant,
{\it i.e.} no flow.
\item
Assume a fixed reaction plane angle; useful for testing analysis code and
debugging.
\item
Assume a randomly fluctuating reaction plane angle which follows a Gaussian
distribution; useful for evaluating reaction plane measurement resolution
effects.
\item
Assume a random reaction plane angle, uniformly distributed between 0$^{\circ}$
and 360$^{\circ}$, which is the way the data will be acquired in the
experiment.
\end{enumerate}

The model assumed for the anisotropic flow is from Ref.~\cite{flow1}, where
\begin{equation}
E\frac{d^3N}{dp^3} = \frac{1}{2\pi p_T} \frac{d^2N}{dp_T dy} \left[1 +
\sum_{n=1}^6 2 V_n(p_T,y) \cos [ n (\phi - \Psi_R) ] \right]~,
\end{equation}
where $\Psi_R$ is the reaction plane angle, $\phi$ is the azimuthal angle
of the outgoing particles, and the parameters $V_n(p_T,y)$ are given by
\cite{flow2},
\begin{eqnarray}
V_n(p_T,y) & = & (V_{n1} + V_{n2} p_T^2) \exp \left( - V_{n3} (y - y_{\rm cm})
^2 \right) ~,~n={\rm even} \\
V_n(p_T,y) & = & (V_{n1} + V_{n2} p_T) {\rm sign}(y - y_{\rm cm})
\left( V_{n3} + V_{n4} \mid\! y - y_{\rm cm} \!\mid^3 \right),n={\rm odd}
\end{eqnarray}
where sign($y$) is $\pm 1$ for $y > 0$ or $y < 0$,
respectively, and $y_{\rm cm}$
is the c.m. rapidity which is 0 for symmetric A+A collisions.  Up to six
Fourier components are permitted; directed flow ($n=1$) and elliptical flow
($n=2$) are most often used.

The user should be aware that careless use of the $V_n(p_T,y)$ parametrization
could have negative impact on the code's performance.  Large values
of $V_{n,i}$ ($i$ = 1,2,3,4) and/or large $p_T$ and $y$ acceptance ranges
can lead to very large, unrealistic anisotropy, particularly near the edges
of the acceptance.  This can result in ``spikey'' $\phi$ dependences that
will cause the random rejection process to perform inefficiently.  The 
$V_n(p_T,y)$ parametrization should therefore be used with caution and the
user should verify that the ``flow'' remains within reasonable bounds
throughout the ($p_T,y$) acceptance range.

\section{Standalone Version}
\label{section:standalone}

The source code and other documentation files are available from the
authors ({\tt ray@physics.utexas.edu}).  Two example input files for the
standalone version of the code have been designed to reproduce the
major particle distributions predicted by {\sc venus} and {\sc vni} for
200~A$\cdot$GeV Au+Au central collisions.  The input
parameters are defined and explained in Appendix~A.
Although resonances may be included, they are not decayed in the
standalone version.

The standalone code produces a log file, {\tt mult\_gen.log},
containing diagnostic information and
error messages if any.  For normal, successful runs this file should be
empty.  An output file, {\tt mult\_gen.out},
is produced which includes a lengthy header containing
the input parameters plus several computed quantities along with the list
of particles.  All events are accumulated in this one file, each 
being separated by event header lines. 
This output file format follows the
{\sc gstar} text format convention described in Ref.~\cite{gstar}.  An
additional, optional output file, {\tt mult\_gen.kin},
can be activated which will write out just
the list of particles with only their {\sc geant} particle ID type,
mass and kinematic information
included.  This is useful for analyses with {\sc paw} since this
ASCII text file can
be easily loaded into {\sc paw}-{\it ntuples}.

\section{Summary}
\label{section:summary}

A fast, convenient Monte Carlo event generator, {\sc mevsim}, has been
developed for STAR simulations.  With this code the user has complete control
over the particle types, the multiplicities and the momentum space distributions
of the particles in the simulated events.  Events generated by {\sc mevsim}
are already in daily use as part of the event reconstruction quality
assurance program in STAR.  {\sc mevsim} events can also be used in many other
types of simulation and physics analysis projects in STAR for years to come.

\clearpage

\begin{table}[h]
\begin{minipage}{\textwidth}
\caption{Resonances included in {\sc mevsim}.}
\label{TableI}
\vspace{8.0pt}
\begin{center}
\begin{tabular}{clll} \hline
Resonance & ID & Mass (GeV)  &  Width (GeV)   \\
\hline
$\rho^+$     &    151   &   0.783   &   0.177     \\
$\rho^-$     &    152   &   0.783   &   0.177     \\
$\rho^0$     &    153   &   0.783   &   0.177     \\
$\omega$     &    154   &   0.782   &   0.0084    \\
$\eta^{\prime}$ & 155   &   0.95750 &   0.000208  \\
$\phi$       &    156   &   1.0194  &   0.00442   \\
J/$\psi$     &    157   &   3.09693 &   0.000068  \\
$\Delta^-$   &    158   &   1.232   &   0.071     \\
$\Delta^0$   &    159   &   1.232   &   0.071     \\
$\Delta^+$   &    160   &   1.232   &   0.071     \\
$\Delta^{++}$&    161   &   1.232   &   0.071     \\
$K^{\ast +}$ &    162   &   0.89183 &   0.0498    \\
$K^{\ast -}$ &    163   &   0.89183 &   0.0498    \\
$K^{\ast 0}$ &    164   &   0.89610 &   0.0505    \\
\hline
\end{tabular}
\end{center}
\end{minipage}
\end{table}

\appendix
\section{Appendix}

This appendix contains a description of the input for the standalone version
of the Monte Carlo event generator,
{\tt multiplicity\_gen.f}

\begin{verbatim}

    (1) n_events - Selected number of events in run. Can be anything
                   .ge. 1.
    (2) n_pid_type - Number of particle ID types to include in the
                     particle list. e.g. pi(+) and pi(-) are counted
                     separately.  The limit is set by parameter npid
                     in the accompanying include file 'Parameter_values.inc'
                     and is presently set at 30.
    (3) model_type - equals 1,2,3,4,5 or 6 so far.  See comments in
                     Function dNdpty to see what is calculated.
                     The models included are:
                   = 1, Factorized mt exponential, Gaussian rapidity model
                   = 2, Pratt non-expanding, spherical thermal source model
                   = 3, Bertsch non-expanding spherical thermal source model
                   = 4, Pratt spherically expanding, thermally equilibrated
                        source model.
                   = 5, Factorized pt and eta distributions input bin-by-bin.
                   = 6, Fully 2D pt,eta distributions input bin-by-bin.
                        NOTE: model_type = 1-4 are functions of (pt,y)
                              model_type = 5,6 are functions of (pt,eta)
    (4) reac_plane_cntrl - Can be either 1,2,3 or 4 where:
                         = 1 to ignore reaction plane and anisotropic flow,
                             all distributions will be azimuthally symm.
                         = 2 to use a fixed reaction plane angle for all
                             events in the run.
                         = 3 to assume a randomly varying reaction plane
                             angle for each event as determined by a
                             Gaussian distribution.
                         = 4 to assume a randomly varying reaction plane
                             for each event in the run as determined by
                             a uniform distribution from 0 to 360 deg.
    (5) PSIr_mean, PSIr_stdev - Reaction plane angle mean and Gaussian
                                std.dev. (both are in degrees) for cases
                                with reac_plane_cntrl = 2 (use mean value)
                                and 3.  Note: these are read in regardless
                                of the value of reac_plane_cntrl.
    (6) MultFac_mean, MultFac_stdev - Overall multiplicity scaling factor
                                      for all PID types; mean and std.dev.;
                                      for trigger fluctuations event-to-evt.
    (7) pt_cut_min,pt_cut_max - Range of transverse momentum in GeV/c.
    (8) eta_cut_min,eta_cut_max - Pseudorapidity range
    (9) phi_cut_min,phi_cut_max - Azimuthal angular range in degrees.
   (10) n_stdev_mult - Number of standard deviations about the mean value
                       of multiplicity to include in the random event-to-
                       event selection process.  The maximum number of
                       steps that can be covered is determined by
                       parameter n_mult_max_steps in the accompanying
                       include file 'Parameter_values.inc' which is
                       presently set at 1000, but the true upper limit for
                       this is n_mult_max_steps - 1 = 999.
   (11) n_stdev_temp - Same, except for the "Temperature" parameter.
   (12) n_stdev_sigma- Same, except for the rapidity width parameter.
   (13) n_stdev_expvel - Same, except for the expansion velocity parameter.
   (14) n_stdev_PSIr   - Same, except for the reaction plane angle
   (15) n_stdev_Vn     - Same, except for the anisotropy coefficients, Vn.
   (16) n_stdev_MultFac - Same, except for the multiplicity scaling factor.
   (17) n_integ_pts - Number of mesh points to use in the random model
                      parameter selection process.  The upper limit is
                      set by parameter nmax_integ in the accompanying
                      include file 'Parameter_values.inc' which is presently
                      set at 100, but the true upper limit for n_integ_pts
                      is nmax_integ - 1 = 99. 
   (18) n_scan_pts  - Number of mesh points to use to scan the (pt,y)
                      dependence of the model distributions looking for
                      the maximum value.  The 2-D grid has
                      n_scan_pts * n_scan_pts points; no limit to size of
                      n_scan_pts.
   (19) irand       - Starting random number seed.

**************************************************************************
    FOR MODEL_TYPE = 1,2,3 or 4:
    Input the following 17 lines for each particle type; repeat these
    set of lines n_pid_type times:

         (a) gpid - Geant Particle ID code number
         (b) mult_mean,mult_variance_control - Mean multiplicity and
                                               variance control where:
             mult_variance_control = 0 for no variance in multiplicity 
             mult_variance_control = 1 to allow Poisson distribution for
                                       particle multiplicities for all events.
             Note that a hard limit exists for the maximum possible
             multiplicity for a given particle type per event.  This is
             determined by parameter factorial_max in accompanying include
             file 'common_facfac.inc' and is presently set at 10000.
         (c) Temp_mean, Temp_stdev - Temperature parameter mean (in GeV)
             and standard deviation (Gaussian distribution assumed).
         (d) sigma_mean, sigma_stdev - Rapidity distribution width (sigma)
             parameter mean and standard deviation (Gaussian distribution
             assumed).
         (e) expvel_mean, expvel_stdev - S. Pratt expansion velocity
             (in units of c) mean and standard deviation (Gaussian 
             distribution assumed).
         (f) Vn_mean(k);  k=1,4  - Anisotropic flow parameters, mean values
                                   for Fourier component n=1.
         (g) Vn_stdev(k); k=1,4  - Anisotropic flow parameters, std.dev.
                                   values for Fourier component n=1.

             Repeat the last two lines of input for remaining Fourier
             components n=2,3...6.  Include all 6 sets of parameters
             even if these are not used by the model for Vn(pt,y) (set
             unused parameter means and std.dev. to 0.0).  List 4 values
             on every line, even though for n=even the 4th quantity is
             not used.

**************************************************************************
    FOR MODEL_TYPE = 5 input the following set of lines for each particle
                       type; repeat these n_pid_type times.

         (a) gpid - Geant Particle ID code number
         (b) mult_mean,mult_variance_control - Mean multiplicity and
                                               variance control where:
             mult_variance_control = 0 for no variance in multiplicity
             mult_variance_control = 1 to allow Poisson distribution for
                                       particle multiplicities for all events.
         (c) pt_start, eta_start - minimum starting values for pt, eta 
                                   input for the bin-by-bin distributions.
         (d) n_pt_bins, n_eta_bins - # input pt and eta bins.
         (e) delta_pt, pt_bin - pt bin size and function value, repeat for
                                each pt bin.
         (f) delta_eta, eta_bin - eta bin size and function value, repeat
                                  for each eta bin.
         (g) Vn_mean(k);  k=1,4  - Anisotropic flow parameters, mean values
                                   for Fourier component n=1.
         (h) Vn_stdev(k); k=1,4  - Anisotropic flow parameters, std.dev.
                                   values for Fourier component n=1.

             Repeat the last two lines of input for remaining Fourier
             components n=2,3...6.  Include all 6 sets of parameters
             even if these are not used by the model for Vn(pt,y) (set
             unused parameter means and std.dev. to 0.0).  List 4 values
             on every line, even though for n=even the 4th quantity is
             not used.

         NOTE: The pt, eta ranges must fully include the requested ranges
               in input #4 and 5 above; else the code execution will stop.

         Also, variable bin sizes are permitted for the input distributions.

         Also, this input distribution is used for all events in the run;
         no fluctuations in this "parent" distribution are allowed from 
         event-to-event.

**************************************************************************
    FOR MODEL_TYPE = 6 input the following set of lines for each particle
                       type; repeat these n_pid_type times.

         (a) gpid - Geant Particle ID code number
         (b) mult_mean,mult_variance_control - Mean multiplicity and
                                               variance control where:
             mult_variance_control = 0 for no variance in multiplicity
             mult_variance_control = 1 to allow Poisson distribution for
                                       particle multiplicities for all events.
         (c) pt_start, eta_start - minimum starting values for pt, eta
                                   input for the bin-by-bin distributions.
         (d) n_pt_bins, n_eta_bins - # input pt and eta bins.
         (e) delta_pt - pt bin size, repeat for each pt bin. 
         (f) delta_eta - eta bin size, repeat for each eta bin.
         (g) i,j,pt_eta_bin(i,j) - read pt (index = i) and eta (index = j)
                                   bin numbers and bin value for full 2D space.
         (h) Vn_mean(k);  k=1,4  - Anisotropic flow parameters, mean values
                                   for Fourier component n=1.
         (i) Vn_stdev(k); k=1,4  - Anisotropic flow parameters, std.dev.
                                   values for Fourier component n=1.

             Repeat the last two lines of input for remaining Fourier
             components n=2,3...6.  Include all 6 sets of parameters
             even if these are not used by the model for Vn(pt,y) (set
             unused parameter means and std.dev. to 0.0).  List 4 values
             on every line, even though for n=even the 4th quantity is
             not used.

         NOTE: The pt, eta ranges must fully include the requested ranges
               in input #4 and 5 above; else the code execution will stop.

         Also, variable bin sizes are permitted for the input distributions.

         Also, this input distribution is used for all events in the run;
         no fluctuations in this "parent" distribution are allowed from
         event-to-event.


\end{verbatim}

\end{document}